\documentclass[aps,prb,twocolumn,superscriptaddress,a4paper]{revtex4-1}
\usepackage[dvipdfmx]{graphicx}

\usepackage{amsmath,amsthm,amssymb}
\usepackage{amsmath,bm}
\usepackage{color}
\usepackage{ifthen}

\usepackage{multirow,bigdelim}

\newcount \commentout
\newcommand{\1}{\mbox{1}\hspace{-0.25em}\mbox{l}}

\def\ket#1{|#1\rangle }
\def\bra#1{\langle #1 |}

\def\II{\mathrm{I\!I}}
\def\I{\mathrm{I}}

\newlength{\figwidth}
\newlength{\figlarge}
\begin{document}
\title{
Square-root topological phase with time-reversal and particle-hole symmetry
}
\author{Tsuneya Yoshida}
\author{Tomonari Mizoguchi}
\author{Yoshihito Kuno}
\author{Yasuhiro Hatsugai}
\affiliation{Department of Physics, University of Tsukuba, Ibaraki 305-8571, Japan}

\date{\today}
\begin{abstract}
Square-root topological phases have been discussed mainly for systems with chiral symmetry. In this paper, we analyze the topology of the squared Hamiltonian for systems preserving the time-reversal and particle-hole symmetry. 
Our analysis elucidates that two-dimensional systems of class CII host helical edge states due to the nontrivial topology of the squared Hamiltonian in contrast to the absence of ordinary topological phases. 
The emergence of the helical edge modes is demonstrated by analyzing a toy model. 
We also show the emergence of surface states induced by the non-trivial topology of the squared Hamiltonian in three dimensions.
\end{abstract}
\maketitle

\section{
Introduction
}
\label{sec: intro}

In these decades, topological aspects of condensed matter systems are extensively studied~\cite{TI_review_Hasan10,TI_review_Qi10}.
A typical example of topological insulators is an integer quantum Hall system~\cite{Klitzing_IQHE_PRL80,Thouless_PRL1982,Halperin_PRB82} where the non-trivial topology in the bulk induces chiral edge modes, i.e., bulk-edge correspondence~\cite{Hatsugai_PRL93}.
The topology in the bulk is enriched for systems preserving Altland-Zirnbauer (AZ) symmetry~\cite{Altland_AZsymm_PRB97,Zrinbauer_AZsymm_2021}, i.e., the time-reversal symmetry~\cite{Kane_2DZ2_PRL05,Kane_Z2TI_PRL05_2}, particle-hole symmetry~\cite{Kitaev_chain_01}, and chiral symmetry~\cite{SSH_PRL79}. 
These variety of topological phases are systematically understood by the ten-fold way classification~\cite{Schnyder_classification_free_2008,Kitaev_classification_free_2009,Ryu_classification_free_2010,Chiu_class_RMP16}; the classification result predicts the presence/absence of topological phases for a given $d$-dimensional system in an AZ symmetry class.
After this progress, topological insulators/superconductors have been extended to various systems; for instance, in these years, higher-order topological insulators~\cite{Hashimoto_HOTI_PRB17,Benalcazar_HOTI_Science17,Benalcazar_HOTI_PRB17,Schindler_HOTI_SciAdv18} and non-Hermitian topological insulators~\cite{Bergholtz_Review19,Yoshida_nHReview_PTEP20,Ashida_nHReview_arXiv20} are actively studied.

Among these extensions, Arkinstall \textit{et al.}~\cite{Arkinstall_SqrtTI_PRB17} proposed square-root topological insulators, which provides a novel insight to topological phases.
They have analyzed a one-dimensional tight-binding model and have demonstrated that the system hosts edge modes induced by the topology of the squared Hamiltonian rather than the original Hamiltonian.
After this proposal, analysis of square-root topological insulators in higher dimensions has been addressed~\cite{Mizoguchi_SqrtHOTI_PRA20,Mizoguchi_SqrtSM_PRB21,Ezawa_SqrtTI_PRR20,Attig_SqrtTI_PRB17,Attig_SqrtMech_PRR19,Dias_2nrtTIarXiv2021,Dias_SqrtTIarXiv2021,Lin_SqrtSkin_OpticsEx2021}, which has elucidated ubiquity of the square-root topological phases. 
For instance, a square-root counterpart of higher-order topological phases are reported by both theoretical~\cite{Mizoguchi_SqrtHOTI_PRA20} and experimental works~\cite{Yan_SqrtPhonon_PRB20,Song_Sqrt-elecir_NanoLett20}. 
In addition, Refs.~\onlinecite{Mizoguchi_SqrtHOTI_PRA20,Ezawa_SqrtTI_PRR20} suggested that for chiral symmetric systems (class AIII) a toy model can be systematically constructed by decorating a lattice~\cite{decol_cites_ftnt}.

Despite of the above progress, square-root topology has not sufficiently explored for systems preserving time-reversal symmetry and particle-hole symmetry~\cite{Symm_Hsq_ftnt}.

In this paper, we analyze systems with time-reversal symmetry and particle-hole symmetry in terms of the square-root perspective. Our analysis elucidates that non-trivial topology of the squared Hamiltonian predicts helical edge modes in two-dimensional systems in class CII [i.e., the systems preserving time-reversal symmetry and the particle-hole symmetry, see Eq.~(\ref{eq: def of TRS})-(\ref{eq: def of CS})] in contrast to the absence of ordinary two-dimensional topological phases in this class.
We also propose a method to construct toy models of square-root topological phases which is a complementary to the method employed in the previous works~\cite{Mizoguchi_SqrtHOTI_PRA20,Ezawa_SqrtTI_PRR20,Mizoguchi_SqrtSM_PRB21}.  Applying our approach to systems of class CII, we demonstrate the emergence of surface Dirac cones for a three-dimensional system as well as the emergence of helical edge modes in a two-dimensional systems.

In Sec.~\ref{sec: CII}, we elucidates that the helical edge modes emerge due to the non-trivial topology of the squared Hamiltonian. 
Applying our approach to construct toy models [see Sec.~\ref{sec: gen const}], we analyze square-root topological phases for other cases of dimensions and symmetry classes in Sec.~\ref{sec: application our approach}. 
A short summary is given in Sec.~\ref{sec: concl}.
Appendices are devoted to the topological invariants of the squared Hamiltonian and technical details.

\section{
Square-root perspective on two-dimensional systems of class CII
}
\label{sec: CII}

We show that square-root topology of two-dimensional systems in class CII induces helical edge modes [see Fig.~\ref{fig: TSC2D_CII}], in contrast to the absence of the ordinary two-dimensional topological phases in this class~\cite{ordinary_TI_table_ftnt}.

\subsection{
Symmetry constraints on the squared Hamiltonian
}
\label{sec: CII symm argument}

Consider a two-dimensional Hamiltonian $H(\bm{k})$ in class CII [see Eq.~(\ref{eq: def of TRS})-(\ref{eq: def of CS})]. 
We show that topology of the squared Hamiltonian $H_{\mathrm{sq}}(\bm{k}):=H^2(\bm{k})$ can be characterized by a $\mathbb{Z}_2$-invariant of two-dimensional systems in class AII.

Let us consider a Hamiltonian $H(\bm{k})$ which preserves time-reversal, particle-hole and chiral symmetry 
\begin{subequations}
\begin{eqnarray}
\label{eq: def of TRS}
T H(\bm{k}) T^{-1}&=& H(-\bm{k}),  \\
\label{eq: def of PHS}
C H(\bm{k}) C^{-1}&=& -H(-\bm{k}),  \\
\label{eq: def of CS}
\Gamma H(\bm{k}) \Gamma^{-1}&=& -H(\bm{k}),
\end{eqnarray}
\end{subequations}
with $\bm{k}:=(k_x,k_y)$ denoting the momentum.
Here, $T$ and $C$ are anti-unitary operators satisfying $[T,C]=0$. 
The unitary operator $\Gamma$ satisfies $\Gamma^2=1$. For symmetry class CII, relations $T^2=C^2=-1$ and $\Gamma=TC$ are satisfied.

The above symmetry constraints result in the following relations
\begin{subequations}
\begin{eqnarray}
\label{eq: TRS on Hsq}
T H_{\mathrm{sq}}(\bm{k}) T^{-1}&=& H_{\mathrm{sq}}(-\bm{k}),  \\
\label{eq: PHS on Hsq}
C H_{\mathrm{sq}}(\bm{k}) C^{-1}&=& H_{\mathrm{sq}}(-\bm{k}),  \\
\label{eq: CS on Hsq}
\Gamma H_{\mathrm{sq}}(\bm{k}) \Gamma^{-1}&=& H_{\mathrm{sq}}(\bm{k}),
\end{eqnarray}
\end{subequations}
with $H_{\mathrm{sq}}(\bm{k}):=H^2(\bm{k})$.

Equation~(\ref{eq: CS on Hsq}) and the relation $\Gamma^2=1$ indicate that $H_{\mathrm{sq}}(\bm{k})$ can be block-diagonalized in the plus and minus sectors of $\Gamma$.
The Hamiltonian of the plus [minus] sector $H_{\mathrm{sq},+}(\bm{k})$ [$H_{\mathrm{sq},-}(\bm{k})$] satisfies
\begin{eqnarray}
T H_{\mathrm{sq},+(-)}(\bm{k}) T^{-1}&=& H_{\mathrm{sq},+(-)}(\bm{k}).
\end{eqnarray}
In addition, in the plus (minus) sector, the operator $C$ is written as $C=-T$ ($C=T$) due to the relations $\Gamma=TC$ and $T^2=-1$.

Therefore, $H_{\mathrm{sq},\pm}$ effectively preserves the time-reversal symmetry; $TH_{\mathrm{sq},\pm}(\bm{k})T^{-1}=H_{\mathrm{sq},\pm}(-\bm{k})$ with $T^2=-1$ which results in quantization of the $\mathbb{Z}_2$-invariant in class AII (see Sec.~\ref{sec: CII Z2 inv}).

\subsection{
Equivalence of the $\mathbb{Z}_2$-invariants of the plus and minus sectors
}
\label{sec: CII Z2 inv}

In the previous section, we have seen that the squared Hamiltonian in each subsector $H_{\mathrm{sq},\pm}$ preserves the time-reversal symmetry with $T^2=-1$. Therefore, the topology of $H_{\mathrm{sq},\pm}$ can be characterized by the $\mathbb{Z}_2$-invariant $\nu=0,1$~(mod~$2$).
In this section, we show that $H_{\mathrm{sq},+}(\bm{k})$ and $H_{\mathrm{sq},-}(\bm{k})$ have the same topology.

\subsubsection{
Brief review of the $\mathbb{Z}_2$-invariant
}
\label{sec: CII Z2 inv review}

Let us start with the definition of $\mathbb{Z}_2$-invariants $\nu_\pm$.
For square-lattice systems, $\nu_\pm$ are defined as~\cite{Kane_2DZ2_PRL05,Fu_2DZ2_PRB06}
\begin{eqnarray}
\label{eq: Z2 inv in 2D}
\nu_{\pm} &=& \frac{1}{2\pi i}\left[ \int^{\pi}_{-\pi} dk_x [A_{\pm,x}(k_x,\pi)-A_{\pm,x}(k_x,0)]  \right. \nonumber \\
          && \quad\quad\quad  \left.-\int^{\pi}_{-\pi} dk_x \int^{0}_{-\pi} dk_y F_{\pm}(\bm{k})\right],
\end{eqnarray}
with $-\pi \leq  k_\mu <\pi$ ($\mu=x,y$). We note that the integral in the last term is taken over a half of the Brillouin zone (BZ).
The Berry connection $A_{\pm,\mu}(\bm{k})$ and the Berry curvature $F(\bm{k})$ are defined as
\begin{subequations}
\begin{eqnarray}
\label{eq: CII A F}
A_{\pm,\mu}(\bm{k}) &=& \sum_{n s} \langle u^s_{\pm n}(\bm{k})|\partial_\mu u^s_{\pm n}(\bm{k})\rangle, \\
F_{\pm}(\bm{k}) &=& \partial_x A_{\pm,y}(\bm{k}) -\partial_y A_{\pm,x}(\bm{k}), 
\end{eqnarray}
\end{subequations}
with $\partial_\mu$ denoting derivative with respect to $k_\mu$.
Here, $\ket{u^s_{\pm n}(\bm{k})}$ denote the eigenstates of $H_{\mathrm{sq},\pm}(\bm{k})$ with $n=1,2,...$ and $s=\I,\II$ which label the energy bands~\cite{I_and_II_ftnt}.
At the time-reversal invariant momenta, eigenstates specified by $(n,\I)$ and $(n,\II)$ form a Kramers pair.

We note that for computation of the $\mathbb{Z}_2$-invariant [Eq.~(\ref{eq: Z2 inv in 2D})], the following gauge should be taken.
\begin{subequations}
\label{eq: gauge constraint}
\begin{eqnarray}
\ket{u^{\I}_{\pm n}(-\bm{k})} &=& T\ket{u^{\II}_{\pm n}(\bm{k})}, \\
\ket{u^{\II}_{\pm n}(-\bm{k})} &=& - T\ket{u^{\I}_{\pm n}(\bm{k})}.
\end{eqnarray}
\end{subequations}

As we see below, 
\begin{eqnarray}
\label{eq: nu+=nu- CII in 2D}
\nu_+ &=& \nu_- \quad (\mathrm{mod}\ 2),
\end{eqnarray}
holds, meaning that $H_{\mathrm{sq},+}(\bm{k})$ and $H_{\mathrm{sq},-}(\bm{k})$ has the same topology.

\subsubsection{
Proof of Eq.~(\ref{eq: nu+=nu- CII in 2D})
}
\label{sec: CII Z2 inv proof}

We show that $H_{\mathrm{sq},+}(\bm{k})$ and $H_{\mathrm{sq},-}(\bm{k})$ have the same topology.

Firstly, we note a relation between $\ket{u^{s}_{+ n}(\bm{k})}$ and $\ket{u^{s}_{- n}(\bm{k})}$.
Consider the case where the energy eigenvalues are non-zero and the Hamiltonian $H(\bm{k})$ is written as~\cite{2n+1x2n+1_ftnt}
\begin{eqnarray}
H(\bm{k})&=& 
\left(
\begin{array}{cc}
0 & Q(\bm{k})\\
Q^\dagger(\bm{k}) & 0
\end{array}
\right),
\end{eqnarray}
with $N\times N$-matrix $Q(\bm{k})$ which maps a state in the minus sector of $\Gamma$ to a state in the plus sector. 
In this basis, the time-reversal operator can be written as 
\begin{eqnarray}
\label{eq: T in chiral basis CII}
T&=& 
\left(
\begin{array}{cc}
U_T & 0 \\
0 & U_T
\end{array}
\right)
\mathcal{K},
\end{eqnarray}
where $U_T$ is the unitary matrix~\cite{UT=-UT_T_ftnt} satisfying $U^T_T=-U_T$, and $\mathcal{K}$ is the operator of complex conjugation.

In this case, the normalized eigenvectors $\ket{u^{s}_{- n}(\bm{k})}$ can be written as
\begin{eqnarray}
\label{eq: u- and u+}
\ket{u^{s}_{-n}(\bm{k})}&=& \frac{1}{\sqrt{ \epsilon_{\mathrm{sq},+sn}(\bm{k}) }} Q^\dagger(\bm{k}) \ket{u^{s}_{+n}(\bm{k})},
\end{eqnarray}
with the normalized eigenvectors $\ket{u^{s}_{+n}(\bm{k})}$. 
Here, eigenvalues of $H(\bm{k})$ are assumed to be non-zero (i.e., eigenvalues $\epsilon_{\mathrm{sq},+sn}(\bm{k})$ of the squared Hamiltonian $H_{\mathrm{sq},+}$ are positive).
In addition, $\ket{u^{s}_{-n}(\bm{k})}$ satisfy the time-reversal constraint [Eq.~(\ref{eq: gauge constraint})], provided that $\ket{u^{s}_{+ n}(\bm{k})}$ satisfy it.

Now, we prove Eq.~(\ref{eq: nu+=nu- CII in 2D}). 
Firstly, we note 
\begin{subequations}
\begin{eqnarray}
\label{eq: CII A- = A+ + f}
A_{-,\mu}(\bm{k}) &=& A_{+,\mu}(\bm{k}) +f_\mu(\bm{k}),
\end{eqnarray}
with
\begin{eqnarray}
f_\mu(\bm{k}) &=& \frac{1}{\sqrt{ \epsilon_{\mathrm{sq},+sn}(\bm{k}) }} \mathrm{tr}[  P_+(\bm{k})  Q(\bm{k})   \nonumber\\
               &&   \quad \quad  \quad \quad  \left( \partial_\mu \frac{1}{\sqrt{ \epsilon_{\mathrm{sq},+sn}(\bm{k}) }} Q^\dagger(\bm{k})\right) ], \\
P_+(\bm{k}) &=& \sum_{ns}\ket{u^{s}_{+n}(\bm{k})} \bra{u^{s}_{+n}(\bm{k})}.
\end{eqnarray}
\end{subequations}
Here, the summation is taken over ``occupied states". (Here, we define ``occupied states" as follows: setting an energy to specify the gap, we can regard the ``occupied states" as eigenstates whose energy is smaller than it.)

By applying Stokes' theorem we can see that the integrals of $f_\mu$ vanishes because $f_\mu(\bm{k})$ is gauge independent.
Therefore, we can see that Eq.~(\ref{eq: nu+=nu- CII in 2D}) holds.

\subsection{
Analysis of a toy model
}
\label{sec: CII toy model}
Following an approach described in Sec.~\ref{sec: gen const} we construct a toy model of the square-root topological phase of class CII in two dimensions. Our analysis demonstrates the presence of helical edge states induced by the square-root topology.

\subsubsection{
Energy spectrum and topological cheracterization
}
\label{sec: CII toy spect}

Suppose that the operators of time-reversal symmetry, particle-hole symmetry, and chiral symmetry are written as
\begin{subequations}
\label{eq: TI CII symm op}
\begin{eqnarray}
T &=& s_2\tau_3\rho_0\mathcal{K},\\
C &=& s_1\tau_2\rho_0\mathcal{K},\\
\Gamma &=& s_3\tau_1\rho_0,
\end{eqnarray}
\end{subequations}
where $s$'s, $\tau$'s, and $\rho$'s are the Pauli matrices.

Here, we show that the following two-dimensional Hamiltonian possesses the square-root topology:
\begin{subequations}
\begin{eqnarray}
\label{eq: model TI CII 2D}
H(\bm{k})&=& H_{\mathrm{D}}(\bm{k}) +m U, \\
H_{\mathrm{D}}(\bm{k}) &=&
p_1(\bm{k}) s_1\tau_0\rho_0
+p_2(\bm{k}) s_3\tau_2\rho_2
+p_3(\bm{k}) s_3\tau_3\rho_2, \nonumber \\
\end{eqnarray}
\end{subequations}
with $\bm{k}=(k_x,k_y)$ denoting the momentum and $U=s_1\tau_1\rho_0$.
Prefactors are defined as $p_1(\bm{k})=2t\sin k_x$, $p_2(\bm{k})=2t\sin k_y$, and $p_3(\bm{k})=2t(\cos k_x+ \cos k_y) -\mu$ with real numbers $t$ and $\mu$.

Noting the commutation relation, $[H_{\mathrm{D}}(\bm{k}),U]=0$, we can compute $H_{\mathrm{sq}}(\bm{k})=H^2(\bm{k})$ as
\begin{eqnarray}
\label{eq: sq model TI CII 2D}
H_{\mathrm{sq}}(\bm{k}) &=& [p^2(\bm{k})+m^2]s_0\tau_0\rho_0 +2m [ p_1(\bm{k}) s_0\tau_1\rho_0  \nonumber \\
&& + p_2(\bm{k}) s_2\tau_3\rho_2 -p_3(\bm{k})s_2\tau_2\rho_2],
\end{eqnarray}
with $p^2(\bm{k})=p^2_1(\bm{k})+p^2_2(\bm{k})+p^2_3(\bm{k})$.

The above Hamiltonian $H_{\mathrm{sq}}(\bm{k})$ can be block-diagonalized with $\Gamma=s_3\tau_1\rho_0$.
The plus sector is spanned by
\begin{subequations}
\label{eq: + basis CII 2D}
\begin{eqnarray}
\ket{+a} &=& 
\frac{1}{\sqrt{2}}
\left(
\begin{array}{c}
1  \\
0
\end{array}
\right)_s
\left(
\begin{array}{c}
1  \\
1
\end{array}
\right)_\tau
\left(
\begin{array}{c}
1  \\
0
\end{array}
\right)_\rho, \\
%
%
\ket{+b} &=& 
\frac{1}{\sqrt{2}}
\left(
\begin{array}{c}
0  \\
1
\end{array}
\right)_s
\left(
\begin{array}{c}
1  \\
-1
\end{array}
\right)_\tau
\left(
\begin{array}{c}
1  \\
0
\end{array}
\right)_\rho, \\
%
%
\ket{+c} &=& 
\frac{1}{\sqrt{2}}
\left(
\begin{array}{c}
1  \\
0
\end{array}
\right)_s
\left(
\begin{array}{c}
1  \\
1
\end{array}
\right)_\tau
\left(
\begin{array}{c}
0  \\
1
\end{array}
\right)_\rho, \\
%
%
\ket{+d} &=& 
\frac{1}{\sqrt{2}}
\left(
\begin{array}{c}
0  \\
1
\end{array}
\right)_s
\left(
\begin{array}{c}
1  \\
-1
\end{array}
\right)_\tau
\left(
\begin{array}{c}
0  \\
1
\end{array}
\right)_\rho.
\end{eqnarray}
\end{subequations}

In this subsector, the Hamiltonian is written as
\begin{eqnarray}
\label{eq: Hsq+ CII}
H_{\mathrm{sq},+}(\bm{k})&=& 
2m[p_1(\bm{k}) \chi_3\rho_0+ p_2(\bm{k}) \chi_2\rho_2 + p_3(\bm{k}) \chi_1\rho_2], \nonumber \\
\end{eqnarray}
where we have omitted the term proportional to the identity matrix [i.e., the first term of Eq.~(\ref{eq: sq model TI CII 2D})].
Matrices $\chi$'s are the Pauli matrices.
The above result is obtained by straight forward calculations, e.g.,
\begin{eqnarray}
s_2\tau_3\rho_2 
\Psi_+
&=&
\Psi_+
\left(
\begin{array}{cccc}
    &   &   & -1 \\
    &   & 1 &  \\
    & 1 &   &  \\
-1  &   &   & 
\end{array}
\right)
:=
\Psi_+
\chi_2\sigma_2,
\end{eqnarray}
with
\begin{eqnarray}
\label{eq: + basis Psi CII 2D}
\Psi_+
&=& 
\left(
\begin{array}{cccc}
\ket{+a} & 
\ket{+b} & 
\ket{+c} & 
\ket{+d}
\end{array}
\right).
\end{eqnarray}

Because $H_{\mathrm{sq},+}(\bm{k})$ preserves the time-reversal symmetry with $T = \chi_2 \sigma_0 \mathcal{K}$ ($T^2=-1$, $T=-C$)~\cite{Tsq+CII_ftnt}, the topological properties can be characterized by the $\mathbb{Z}_2$-invariant [see Eq.~(\ref{eq: Z2 inv in 2D})].
In the presence of the inversion symmetry, the $\mathbb{Z}_2$-invariant $\nu_+$ can be computed from the parity of ``occupied states" at the time-reversal invariant momenta~\cite{Fu_Z2TI_inv_PRB07}. 
Noting that $H_{\mathrm{sq},+}(\bm{k})$ satisfies $P H_{\mathrm{sq},+}(\bm{k}) P^{-1} = H_{\mathrm{sq},+}(-\bm{k})$ with $P=\chi_1\rho_2$, we can see that 
among the time-reversal invariant momenta $(0,0)$, $(\pi,0)$, $(0,\pi)$, and $(\pi,\pi)$, the parity eigenvalue of the ``occupied band" is plus only at $(0,0)$. Because the product of the parity eigenvalues at the time-reversal invariant momenta is minus~\cite{Z2_parity_ftnt}, we have $\nu_+=1$.

\begin{figure}[!t]
\begin{minipage}{0.7\hsize}
\begin{center}
\includegraphics[width=1\hsize,clip]{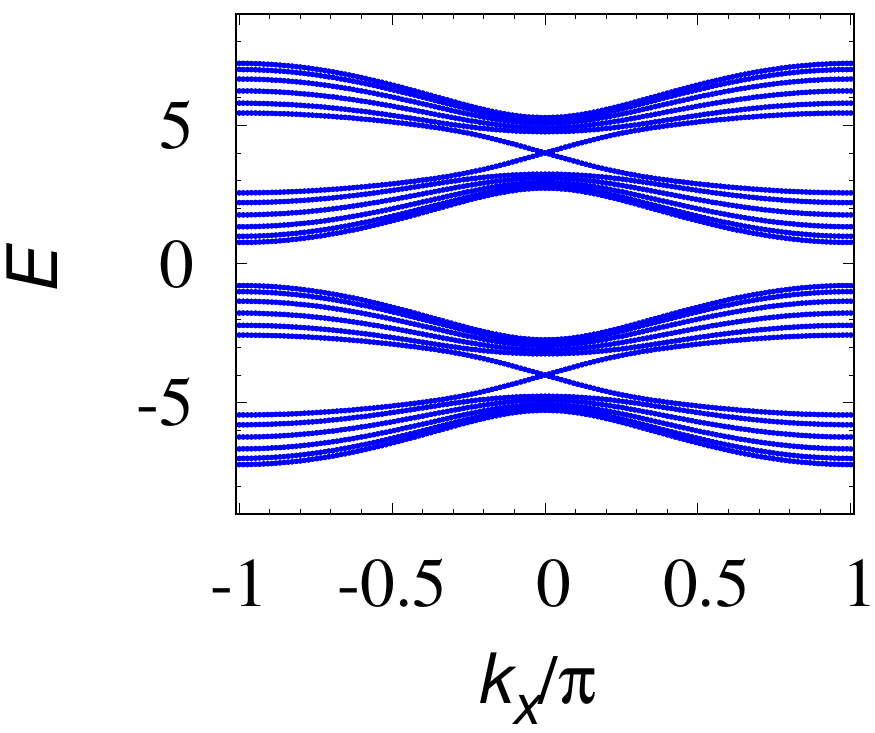}
\end{center}
\end{minipage}
\caption{
Energy spectrum for $t=0.5$, $\mu=1.3$, and $m=4$.
These data are obtained by imposing the periodic (open) boundary condition along the $x$- ($y$-) direction.
Along the $y$-direction $L_y=6$ sites are aligned.
}
\label{fig: TSC2D_CII}
\end{figure}

The square-root topology of the Hamiltonian~(\ref{eq: model TI CII 2D}) characterized by $\nu_+=1$ predicts edge modes. 
In Fig.~\ref{fig: TSC2D_CII}, we plot the band structure obtained under the cylinder geometry [i.e., the periodic (open) boundary condition is imposed along the $x$- ($y$-) direction].
In this figure, we can find the helical edge modes around $E\sim\pm4$.

The above results elucidate that for two-dimensional systems of class CII, helical edge modes emerge due to non-trivial topology of the squared Hamiltonian despite the ordinary two-dimensional topological phases are absent in this class.

\section{
Approach to construct toy models 
}
\label{sec: gen const}

We describe our approach to construct a toy model of a square-root topological phase which is complementary to the method employed in previous works~\cite{Mizoguchi_SqrtHOTI_PRA20,Ezawa_SqrtTI_PRR20,Mizoguchi_SqrtSM_PRB21}.
Our approach is available when the following two conditions are satisfied.
(i) There exists a topological phase described by a Hamiltonian which preserves an additional symmetry [see Eq.~(\ref{eq: symm on U})].
(ii) The presence/absence of the additional symmetry does not affect the topology of the squared Hamiltonian.

\subsection{
Generic framework
}
\label{sec: gen const detail}
When the presence/absence of the additional symmetry does not affect the topology of the squared Hamiltonian, a Hamiltonian with the additional symmetry serves a toy model.
This is because topology of the squared Hamiltonian is equivalent to the topology of the original Hamiltonian.

In the following, we show that topology of the squared Hamiltonian is equivalent to the topology of the original Hamiltonian in the presence of the additional symmetry.
Consider a Hamiltonian in a given symmetry class (e.g., class CII) whose squared Hamiltonian with additional symmetry is topologically nontrivial.
Here, the additional symmetry is described by the commutation relation of the squared Hamiltonian and a unitary matrix $U$ [$U^2=\1$] which also preserves the AZ symmetry; for instance in the case of class CII, $U$ satisfies
\begin{subequations}
\label{eq: symm on U}
\begin{eqnarray}
T U T^{-1} &=& U, \\
C U C^{-1} &=& -U, \\
\Gamma U \Gamma^{-1} &=& -U.
\end{eqnarray}
\end{subequations}
By making use of the adiabatic continuation, we can deform the squared Hamiltonian to
\begin{eqnarray}
\label{eq: simplified Hsq gen}
H'_{\mathrm{sq}}(\bm{k})&=& \sum_{i=1,...,d+1} p_i(\bm{k}) \tilde{\gamma}_i,
\end{eqnarray}
where $\tilde{\gamma}$'s are Hermitian matrix satisfying $\{ \tilde{\gamma}_i , \tilde{\gamma}_j\}=2\delta_{ij}$ for $i,j=1,...,d+1$ and $[\tilde{\gamma}_i,U]=0$ for $i=1,...,d+1$.
Here, $p(\bm{k})$'s are scalars which are chosen so that $H'_{\mathrm{sq}}(\bm{k})$ is reduced to the gapped Dirac Hamiltonian [for instance see, just below of Eq.~(\ref{eq: model TI CII 2D})].

Thanks to the additional symmetry, we can find the square-root Hamiltonian $H(\bm{k})$ of the above simplified model.
Namely, the square-root Hamiltonian, which preserves the symmetry constraints, is written as
\begin{subequations}
\label{eq: gen sqrtDirac}
\begin{eqnarray}
H(\bm{k})&=& H_\mathrm{D}(\bm{k}) +m U, \\
H_\mathrm{D}(\bm{k})&=& \sum_{i=1,...,d+1} p_i(\bm{k}) \gamma_i,
\end{eqnarray}
\end{subequations}
where the matrices $\gamma_i=U\tilde{\gamma_i}$ satisfy $\{ \gamma_i , \gamma_j\}=2\delta_{ij}$ for $i,j=1,...,d+1$.
This fact can be seen by a straightforward calculation:
\begin{eqnarray}
H^2(\bm{k})&=&  (p^2(\bm{k})+m^2)\1 +2mU H_\mathrm{D}(\bm{k}),
\end{eqnarray}
%
with $p^2(\bm{k})=\sum_{i=1,...d+1}p^2_i(\bm{k})$ and $\1$ being the identity matrix.
The above result indicates that in the presence of the additional symmetry [see Eq.~(\ref{eq: symm on U})], the topology of the squared Hamiltonian is identical to the topology of the original Hamiltonian.

Therefore, we can construct the toy model in a given symmetry class by obtaining $H_{\mathrm{D}}(\bm{k})$ in the following steps.
(i) Define the operators of the given symmetry class.
(ii) Introduce $U$ satisfying Eq.~(\ref{eq: symm on U}).
(iii) Prepare the gapped Dirac Hamiltonian $H_{\mathrm{D}}(\bm{k})$ so that $H_{\mathrm{D}}(\bm{k})$ preserves the additional symmetry as well as the AZ symmetry.

\subsection{
A remark on topology of two-dimensional systems in class CII
}
We note that for class CII, the presence of the additional symmetry does not change the topological properties.
As we have seen in Sec.~\ref{sec: CII symm argument}, the topology of the squared Hamiltonian is described by $H_{\mathrm{sq},+}$ and $H_{\mathrm{sq},-}$ which preserve time-reversal symmetry.
Equation~(\ref{eq: symm on U}) indicates that no further constraint is imposed on $H_{\mathrm{sq},\pm}$. The additional symmetry just requires that the topology of $H_{\mathrm{sq},+}$ and that of $H_{\mathrm{sq},-}$ are the same, which is satisfied even without the additional symmetry (see Sec.~\ref{sec: CII Z2 inv proof}).
Therefore, our approach can be applied to construct a two-dimensional model of class CII [see Sec.~\ref{sec: CII toy spect}].

\section{
Application to other symmetry classes
}
\label{sec: application our approach} 

As we have seen in Sec.~\ref{sec: gen const detail}, a toy model of a given symmetry class can be constructed by considering the system with the additional symmetry when the additional symmetry does not change the topology.
In other words, a necessary condition for our approach to be applicable is that for a given symmetry class and dimensions, there exists a topological phase in a system with the additional symmetry~\cite{ness_cond_ftnt}.

In Sec.~\ref{sec: topo class results}, we elucidate when a topological phase exists in the presence of the additional symmetry by addressing the corresponding topological classification. 
The obtained classification results help us to find a toy model showing square-root topology in two dimensions for class AIII [see Sec.~\ref{sec: AIII in 2D}], and in three dimensions in class CII [see Sec.~\ref{sec: CII in 3D}].

\subsection{
Topological classifications of systems with the additional symmetry
}
\label{sec: topo class results}

The topological classification for systems with the additional symmetry can be carried out by analyzing the symmetry of the block-diagonalized Hamiltonian with the unitary matrix $U$ ($U^2=1$).
Classification results of topological phases with the additional symmetry are summarized in Table~\ref{table: class_table_sqrt-TI}.
\begin{table}[htb]
\begin{center}
\begin{tabular}{ c c c c c c c} \hline\hline
class of $H(\bm{k})$ & $T$ & $C$ & $\Gamma$ & $d=1$           & $d=2$           & $d=3$            \\ \hline
A                    &  0  & 0   & 0  & 0               & $\mathbb{Z}$    &  0               \\ 
AIII                 &  0  & 0   & 1  & 0               & $\mathbb{Z}$    &  0               \\ \hline
AI                   &  1  & 0   & 0  & 0               & 0               &  0               \\ 
BDI                  &  1  & 1   & 1  & 0               & 0               &  0               \\ 
D                    &  0  & 1   & 0  & 0               & $\mathbb{Z}$    &  0               \\ 
DIII                 &  -1 & 1   & 1  & 0               &  $\mathbb{Z}_2$ &  $\mathbb{Z}_2$  \\ 
AII                  &  -1 & 0   & 0  & 0               &  $\mathbb{Z}_2$ &  $\mathbb{Z}_2$  \\ 
CII                  &  -1 & -1  & 1  & 0               &  $\mathbb{Z}_2$ &  $\mathbb{Z}_2$  \\ 
C                    &  0  & -1  & 0  & 0               & $\mathbb{Z}$    &  0               \\
CI                   & 1   & -1  & 1  & 0               & 0               &  0  \\  \hline\hline
\end{tabular}
\end{center}
\caption{
Classification results of $d$-dimensional topological insulators/superconductors with the additional symmetry [see Eqs.~(\ref{eq: def of TRS})-(\ref{eq: def of CS})].
The second, the third, and the fourth columns specify a type of the corresponding operators; $\pm 1$ in the second [the third] column represents the sign of $T^2=\pm 1$ [$C^2=\pm 1$]. (``$0$" denotes that the corresponding symmetry is absent.)
We note that the absence of chiral symmetry or particle-hole symmetry allows $U$ to be the identity matrix. Thus, topological insulators in class A, AI, AII are ordinary topological insulators.
}
\label{table: class_table_sqrt-TI}
\end{table}

As explained in Appendix~\ref{sec: classfication app}, the classification results are the same for each of the following three groups: (i) symmetry classes A, AIII, D, and C; (ii) symmetry classes AI, BDI, and CI; (iii) symmetry classes AII, DIII, and CII.
This is because the particle-hole symmetry and chiral symmetry are not closed in the subsector of $U$.

Table~\ref{table: class_table_sqrt-TI} indicates that in the presence of the additional symmetry, there exists a topological phase for a two-dimensional system in class AIII and for a three-dimensional system in class CII.

By making use of these results, we demonstrate the presence of the square-root topological phases whose topology is maintained even in the absence of the additional symmetry.

\subsection{
Two-dimensional system in class AIII
}
\label{sec: AIII in 2D}

Consider a Hamiltonian $H(\bm{k})$ for a two-dimensional system with chiral symmetry [see Eq.~(\ref{eq: def of CS})].
As mentioned in Sec.~\ref{sec: CII symm argument}, the squared Hamiltonian $H_{\mathrm{sq}}(\bm{k})$ can be block-diagonalized with subsectors of the chiral operator $\Gamma$ ($\Gamma^2=1$).

The squared Hamiltonian $H_{\mathrm{sq},\pm}(\bm{k})$ of the plus (minus) sector preserves no symmetry and its topology is characterized by the Chern number [see Eq.~(\ref{eq: def of Chern})].
As explained in Appendix~\ref{sec: AIII 2D app}, the Chern number for the plus sector should be equal to the Chern number for the minus sector. 
In addition, the presence/absence the additional symmetry does not affect the topology (see Appendix~\ref{sec: AIII 2D app}).

Therefore, our approach is available for two-dimensional systems in class AIII.
In the following, we see that a toy model indeed hosts a chiral edge state induced by the square-root topology.

\subsubsection{
Edge modes and the topological characterization
}
\label{sec: AIII in 2D edge}
Let us consider a Hamiltonian of $4\times 4$-matrix.
The operator of chiral symmetry is given by
\begin{eqnarray}
\Gamma &=& s_0\tau_3.
\end{eqnarray}
The operator $U$ satisfying Eq.~(\ref{eq: symm on U}c) can be chosen as $U=\sigma_0\tau_1$.

In this case, we can construct the following Hamiltonian preserving the symmetry
\begin{subequations}
\begin{eqnarray}
\label{eq: model AIII 2D}
H(\bm{k})&=& H_{\mathrm{D}}(\bm{k}) + m \sigma_0\tau_1, \\ 
H_{\mathrm{D}}(\bm{k}) &=& p_1(\bm{k}) \sigma_1\tau_1+ p_2(\bm{k}) \sigma_2\tau_1+p_3(\bm{k}) \sigma_3\tau_1, \nonumber \\
\end{eqnarray}
\end{subequations}
where $\bm{k}=(k_x,k_y)$ denotes the momentum. Prefactors $p_i$ ($i=1,2,3$) are defined just below Eq.~(\ref{eq: model TI CII 2D}).

\begin{figure}[!t]
\begin{minipage}{0.47\hsize}
\begin{center}
\includegraphics[width=1\hsize,clip]{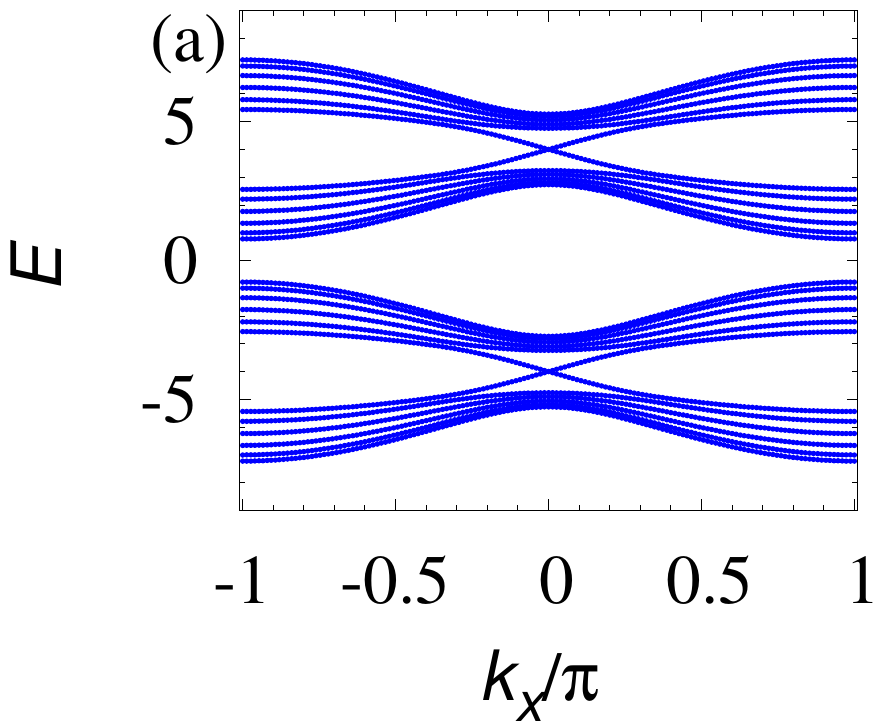}
\end{center}
\end{minipage}
\begin{minipage}{0.47\hsize}
\begin{center}
\includegraphics[width=1\hsize,clip]{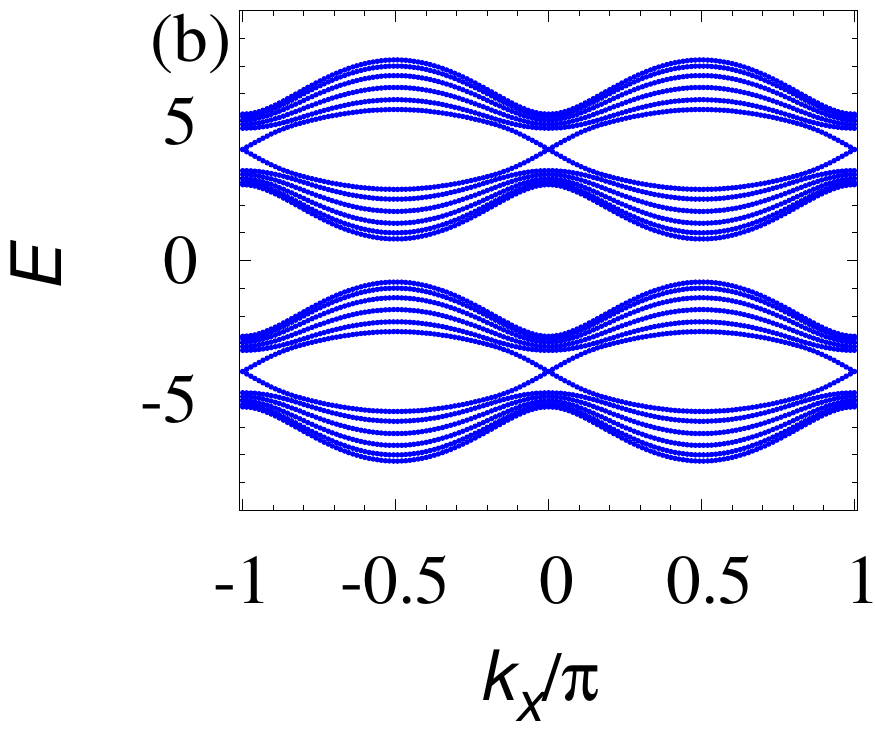}
\end{center}
\end{minipage}
\caption{
Energy spectrum obtained under the cylinder geometry.
Panel (a) is obtained for $p_1(\bm{k})=2t\sin k_x$, $p_2(\bm{k})=2t\sin k_y$, and $p_3(\bm{k})=2t(\cos k_x +\cos k_y) -\mu$.
Panel (b) is obtained for $p_1(\bm{k})=2t\sin 2k_x$, $p_2(\bm{k})=2t\sin k_y$, and $p_3(\bm{k})=2t(\cos 2k_x +\cos k_y) -\mu$. In this case, the Chern number takes two $N_{\mathrm{Ch},+}=2$ inducing two chiral edge modes.
These data are obtained for $t=0.5$, $\mu=1.3$, and $m=4$.
}
\label{fig: TI2D_AIII}
\end{figure}

The squared Hamiltonian can be block-diagonalized with the matrix $\Gamma=s_0\tau_3$. The Hamiltonian in the plus sector is written as
\begin{eqnarray}
H_{\mathrm{sq},+}(\bm{k})&=& [p^2(\bm{k})+m^2]\sigma_0 \nonumber \\
                          &&+2m [p_1(\bm{k}) \sigma_1+ p_2(\bm{k}) \sigma_2+p_3(\bm{k}) \sigma_3]. 
\end{eqnarray}
We can see that the Chern number $N_{\mathrm{Ch},+}$ [for the definition see Eq.~(\ref{eq: def of Chern})] takes one, predicting the presence of a chiral edge mode.

Correspondingly, the numerical data under the cylinder geometry indicate the presence of the chiral edge mode [see Fig.~\ref{fig: TI2D_AIII}(a)].

The above results elucidate that the model defined in Eq.~(\ref{eq: model AIII 2D}) hosts chiral edge modes due to the square-root topology characterized by the Chern number.
Reference~\onlinecite{Ezawa_SqrtTI_PRR20} has constructed a model whose topology is essentially the same as Eq.~(\ref{eq: model AIII 2D}). 
We note however that our approach different from the one used in previous works~\cite{{Mizoguchi_SqrtHOTI_PRA20,Ezawa_SqrtTI_PRR20,Mizoguchi_SqrtSM_PRB21}}. 
While the previous approach~\cite{{Mizoguchi_SqrtHOTI_PRA20,Ezawa_SqrtTI_PRR20,Mizoguchi_SqrtSM_PRB21}} is based on the real-space picture, our approach is based on the momentum-space picture, which allows us to construct a model with $N_{\mathrm{Ch},+}=2$ [see Fig.~\ref{fig: TI2D_AIII}(b)].

\subsection{
Three-dimensional system in class CII
}
\label{sec: CII in 3D}
Consider a Hamiltonian $H(\bm{k})$ for a three-dimensional system of class CII [see Eqs.~(\ref{eq: def of TRS})-(\ref{eq: def of CS})].
As mentioned in Sec.~\ref{sec: CII symm argument}, the squared Hamiltonian $H_{\mathrm{sq}}(\bm{k})$ can be block-diagonalized with subsectors of the chiral operator $\Gamma$ ($\Gamma^2=1$).

The squared Hamiltonian $H_{\mathrm{sq},\pm}(\bm{k})$ of the plus (minus) sector preserves the time-reversal symmetry with $T^2=-1$. Therefore, the topology of $H_{\mathrm{sq},\pm}(\bm{k})$ can be characterized by the $\mathbb{Z}_2$-invariant for three-dimensional systems.
As proven in Appendix~\ref{sec: CII 3D app} topology of the plus sector is the same as topology of the minus sector.
In addition, the presence/absence of the additional symmetry does not change the topology (Appendix~\ref{sec: CII 3D app}).

Therefore, our approach is available for three-dimensional systems in class CII.
In the following, we see that a toy model indeed hosts surface states induced by the square-root topology.

\subsubsection{
Surface modes and the topological characterization
}
\label{sec: CII in 3D surf}

Consider the following thee-dimensional Hamiltonian,
\begin{subequations}
\begin{eqnarray}
\label{eq: model TI CII 3D}
H(\bm{k})&=& H_{\mathrm{D}}(\bm{k}) +m s_1\tau_1\rho_0,\\ 
H_{\mathrm{D}}(\bm{k}) &=& p_1(\bm{k}) s_3\tau_3\rho_1
+p_2(\bm{k}) s_3\tau_3\rho_3\nonumber \\
&&+p_3(\bm{k}) s_1\tau_0\rho_0
+p_4(\bm{k}) s_3\tau_3\rho_2,
\end{eqnarray}
\end{subequations}
where $\bm{k}=(k_x,k_y,k_z)$ denotes the momentum.
Prefactors are defined as $p_1(\bm{k})=2t\sin k_x$, $p_2(\bm{k})=2t\sin k_y$, $p_3(\bm{k})=2t\sin k_z$, and $p_4(\bm{k})=2t(\cos k_x+ \cos k_y + \cos k_z) -\mu$.
This system preserves the time-reversal and particle-hole symmetry whose operator are defined in Eq.~(\ref{eq: TI CII symm op}).

\begin{figure}[!t]
\begin{minipage}{0.45\hsize}
\begin{center}
\includegraphics[width=1\hsize,clip]{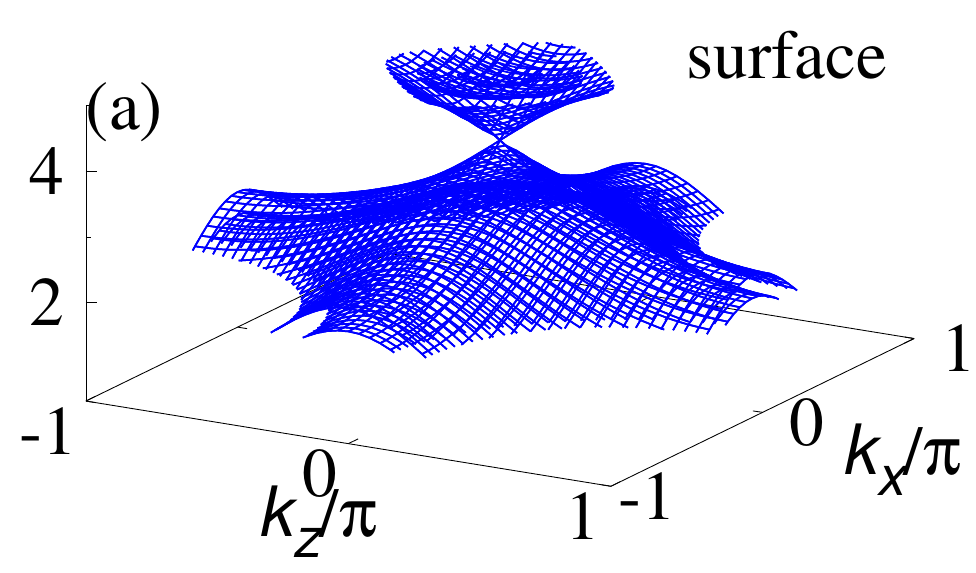}
\end{center}
\end{minipage}
\begin{minipage}{0.45\hsize}
\begin{center}
\includegraphics[width=1\hsize,clip]{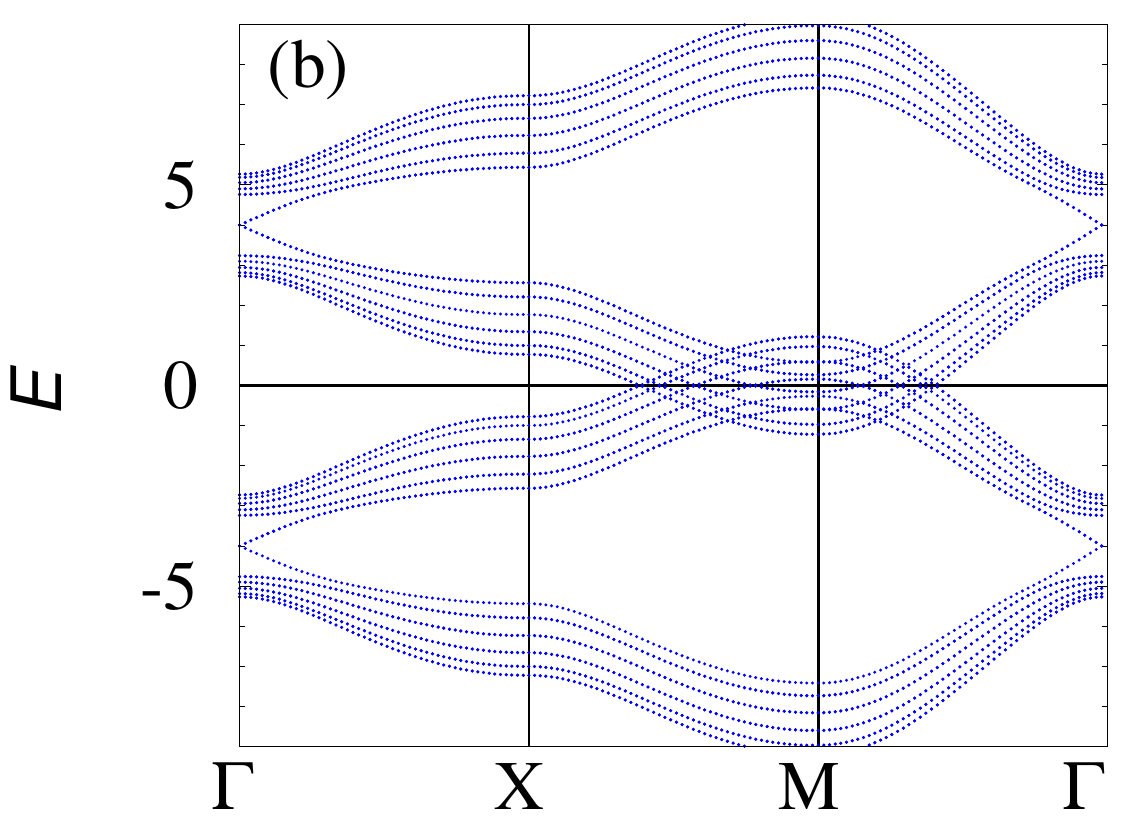}
\end{center}
\end{minipage}
\caption{
(a): The surface band structure around $E=4$.
(b): The surface band structure along the high symmetry points, $\Gamma$ [$(0,0)$], $X$ [$(\pi,0)$], and $M$ [$(\pi,\pi)$].
These data are obtained for  $t=0.5$, $\mu=2.3$, $m=4$. 
Here, the periodic boundary condition is imposed for the $x$- and $z$-direction, while the open boundary condition is imposed for the $y$-direction.
Along the $y$-direction, $L_y=6$ sites are aligned. 
}
\label{fig: TSC3D_CII}
\end{figure}

The squared Hamiltonian $H_{\mathrm{sq}}(\bm{k})$ can be block-diagonalized with the matrix $\Gamma=s_3\tau_1\rho_0$.
The Hamiltonian in the plus sector is written as
\begin{eqnarray}
H_{\mathrm{sq},+}(\bm{k})&=& [p^2(\bm{k})+m^2]\chi_0\rho_0 +2m[p_1(\bm{k})\chi_1\rho_1+p_2(\bm{k})\chi_1\rho_3 \nonumber \\
                          && +p_3(\bm{k})\chi_3\rho_0+p_4(\bm{k})\chi_1\rho_2].
\end{eqnarray}
Here, the plus sector is spanned by a set of the vectors introduced in Sec.~\ref{sec: CII toy model} [see Eqs.~(\ref{eq: + basis CII 2D})~and~(\ref{eq: + basis Psi CII 2D})].

As is the case of two-dimensional system [see Sec.\ref{sec: CII toy model}], the $\mathbb{Z}_2$-invariant $\nu^{\mathrm{3D}}_+$ [for the definition see Eq.~(\ref{eq: def of 3D Z2 app})] can be computed from the parity eigenvalues of ``occupied states" at the time-reversal invariant momenta when the system is inversion symmetric.
We note that $H_{\mathrm{sq},+}(\bm{k})$ preserves the inversion symmetry with $P=\chi_1\rho_2$. Computing the product of the parity eigenvalues at the time-reversal invariant momenta, we can see that the $\mathbb{Z}_2$-invariant $\nu^{\mathrm{3D}}_+$ takes one, predicting the presence of surface states.

Correspondingly, Dirac cones can be found around $E\sim\pm 4$ in Figs.~\ref{fig: TSC3D_CII}(a)~and~\ref{fig: TSC3D_CII}(b) which are obtained by imposing the periodic (open) boundary condition along the $x$- and $z$-directions (the $y$-direction).

The above results elucidate that the model defined in Eq.~(\ref{eq: model TI CII 3D}) hosts Dirac surface states due to the square-root topology characterized by the $\mathbb{Z}_2$-invariant $\nu^{\mathrm{3D}}_{+}$.

\section{
Conclusion
}
\label{sec: concl}
In this paper, we have analyzed topology of the squared Hamiltonian for systems preserving time-reversal and particle-hole symmetry. 
Our analysis elucidates that nontrivial topology of the squared Hamiltonian induces helical edge modes at the boundary of the two-dimensional system in class CII in contrast to the absence of the ordinary two-dimensional topological phases in this class.

We have also proposed a method to construct toy models which is complementary to the previous one. Based on our approach, we demonstrate the emergence of helical edge modes in two-dimensional systems of class CII as well as chiral edge modes in two-dimensional systems in class AIII and surface Dirac cones in three-dimensional systems in class CII.

\section*{
Acknowledgments
}
T.Y. thanks Takahiro Fukui for fruitful comments on the $\mathbb{Z}_2$-invariant for time-reversal symmetric systems.
This work is supported by JPSP Grant-in-Aid for Scientific Research on Innovative Areas ``Discrete Geometric Analysis for Materials Design”: Grants No. JP20H04627 (T.Y.).  This work is also supported by the JSPS KAKENHI, Grants No.~JP17H06138, No.~JP20K14371 (T.M.).


\begin{thebibliography}{50}
\bibitem [{dec()}]{decol_cites_ftnt}%
  \BibitemOpen
  \href@noop {} {}\bibinfo {note} {{ This approach is introduced in
  Ref.~\onlinecite{Mizoguchi_SqrtHOTI_PRA20} and is employed to construct toy
  models of square-root topological insulators in one and two dimensions in
  Ref.~\onlinecite{Ezawa_SqrtTI_PRR20} }}\BibitemShut {NoStop}%
\bibitem [{Sym()}]{Symm_Hsq_ftnt}%
  \BibitemOpen
  \href@noop {} {}\bibinfo {note} {{ While Ref.~\onlinecite{Ezawa_SqrtTI_PRR20}
  has analyzed a one-dimensional system whose squared Hamiltonian becomes
  Kitaev chain, the physical origin of symmetry of squared Hamiltonian is not
  clear }}\BibitemShut {NoStop}%
\bibitem [{ord()}]{ordinary_TI_table_ftnt}%
  \BibitemOpen
  \href@noop {} {}\bibinfo {note} {{ The absence of the ordinary
  topological phase can be seen in Table~\ref{table: class_table_ordinary-TI}
  in Appendix~\ref{sec: classfication app} }}\BibitemShut {NoStop}%
\bibitem [{I_a()}]{I_and_II_ftnt}%
  \BibitemOpen
  \href@noop {} {}\bibinfo {note} {{ When the $z$-comportment of
  spins is a conserved quantity, $s=\I$ ($s=\II$) labels the up-spin
  (down-spin) state }}\BibitemShut {NoStop}%
\bibitem [{2n+()}]{2n+1x2n+1_ftnt}%
  \BibitemOpen
  \href@noop {} {}\bibinfo {note} {{ We consider a Hamiltonian of a $2N\times
  2N$-matrix. The case of $2(N+1) \times 2(N+1)$-matrix can be discussed by
  adding a trivial band which does not change the topology }}\BibitemShut
  {NoStop}%
\bibitem [{UT=()}]{UT=-UT_T_ftnt}%
  \BibitemOpen
  \href@noop {} {}\bibinfo {note} {{ Because $T^2=-1$ holds, we have
  $U_TU^*_T=-1$. Recalling that $U_T$ is a unitary matrix, we obtain $U_T
  =-U^T_T$ }}\BibitemShut {NoStop}%
\bibitem [{Tsq()}]{Tsq+CII_ftnt}%
  \BibitemOpen
  \href@noop {} {}\bibinfo {note} {{ The time-reversal operator for
  the plus sector is obtained in a similar way as Eq.~(\ref{eq: Hsq+ CII})
  }}\BibitemShut {NoStop}%
\bibitem [{Z2_()}]{Z2_parity_ftnt}%
  \BibitemOpen
  \href@noop {} {}\bibinfo {note} {{ We can observe Kramers degeneracy at the
  time-reversal invariant momenta. The $\mathbb{Z}_2$-invariant is
  obtained~\cite{Fu_Z2TI_inv_PRB07} by computing the product of parity
  eigenvalues of only ``occupied states" labeled with $s=\I$ }}\BibitemShut
  {NoStop}%
\bibitem [{nes()}]{ness_cond_ftnt}%
  \BibitemOpen
  \href@noop {} {}\bibinfo {note} {{ For instance, if the topological
  phases are absent for $d$-dimensional systems of a given symmetry class
  preserving the additional symmetry, our approach is not available
  }}\BibitemShut {NoStop}%
\bibitem [{hsq()}]{hsq_ftnt}%
  \BibitemOpen
  \href@noop {} {}\bibinfo {note} {{ We note that $h_{\mathrm{sq},\pm}(\bm{k})$
  differs from $H_{\mathrm{sq},\pm}(\bm{k})$. The former (latter) is the
  block-diagonalized Hamiltonian with $U$ ($\Gamma$) }}\BibitemShut {NoStop}%
\end{thebibliography}
%


\appendix
\section{
Proof of Eq.~(\ref{eq: u- and u+})
}
\label{sec: proof of u- and u+}

Firstly, we note the relations $H_{\mathrm{sq},+}(\bm{k})=Q(\bm{k})Q^\dagger(\bm{k})$ and $H_{\mathrm{sq},-}(\bm{k})=Q^\dagger(\bm{k})Q(\bm{k})$ which can be seen with the following calculation:
\begin{eqnarray}
H_{\mathrm{sq}}&=& 
\left(
\begin{array}{cc}
Q(\bm{k})Q^\dagger(\bm{k}) & 0 \\
0 & Q^\dagger(\bm{k})Q(\bm{k})
\end{array}
\right).
\end{eqnarray}

The above relations indicate that when $\ket{u^{s}_{+n}(\bm{k})}$ are eigenvectors of $H_{\mathrm{sq},+}(\bm{k})$,
$Q^\dagger(\bm{k}) \ket{u^{s}_{+n}(\bm{k})}$ are eigenvectors of $H_{\mathrm{sq},-}(\bm{k})$, which can be seen as follows.
For given eigenstates $\ket{u^{s}_{+n}(\bm{k})}$
\begin{eqnarray}
H_{\mathrm{sq},+}(\bm{k}) \ket{u^{s}_{+ n}(\bm{k})} &=& \ket{u^{s}_{+ n}(\bm{k})}\epsilon_{\mathrm{sq},+sn}(\bm{k}),
\end{eqnarray}
with eigenvalues $\epsilon_{\mathrm{sq},+sn}(\bm{k})$, we have~\cite{Ezawa_SqrtTI_PRR20}
\begin{eqnarray}
Q^\dagger(\bm{k}) H_{\mathrm{sq},+}(\bm{k}) \ket{u^{s}_{+ n}(\bm{k})} &=& Q^\dagger(\bm{k}) \ket{u^{s}_{+ n}(\bm{k})}\epsilon_{\mathrm{sq},+sn}(\bm{k}) \nonumber\\
\Leftrightarrow H_{\mathrm{sq},-}(\bm{k})Q^\dagger(\bm{k}) \ket{u^{s}_{+ n}(\bm{k})} &=& Q^\dagger(\bm{k}) \ket{u^{s}_{+ n}(\bm{k})}\epsilon_{\mathrm{sq},+sn}(\bm{k}).\nonumber \\
\end{eqnarray}
Therefore, $Q(\bm{k}) \ket{u^{s}_{+ n}(\bm{k})}$ are eigenvectors of $H_{\mathrm{sq},-}(\bm{k})$.

Assuming that eigenvalues of $H(\bm{k})$ are non-zero [i.e., $\epsilon_{\mathrm{sq},+sn}(\bm{k})$ are positive], the normalized eigenvectors are written as
\begin{eqnarray}
\ket{u^{s}_{-n}(\bm{k})}&=& \frac{1}{\sqrt{ \epsilon_{\mathrm{sq},+sn}(\bm{k}) }} Q^\dagger(\bm{k}) \ket{u^{s}_{+n}(\bm{k})},
\end{eqnarray}
with the normalized eigenvectors $\ket{u^{s}_{+n}(\bm{k})}$.

In addition, $\ket{u^{s}_{-n}(\bm{k})}$ satisfy the time-reversal constraint [Eq.~(\ref{eq: gauge constraint})], provided that $\ket{u^{s}_{+ n}(\bm{k})}$ satisfy it.
This can be seen by noting the relation
\begin{eqnarray}
\label{eq: U_T Q^*(k) U_T=Q(-k)}
U_T Q^*(\bm{k})U^\dagger_T&=&Q(-\bm{k}),
\end{eqnarray}
which holds because $H(\bm{k})$ preserves the time-reversal symmetry [see Eqs.~(\ref{eq: def of TRS})~and~(\ref{eq: T in chiral basis CII})].
Namely, applying $T$, we have
\begin{eqnarray}
T\ket{u^{s}_{-n}(\bm{k})} &=&\frac{1}{\sqrt{ \epsilon_{\mathrm{sq},+sn}(\bm{k}) }} U_T Q^T(\bm{k}) \mathcal{K} \ket{u^{s}_{+n}(\bm{k})} \nonumber \\
                         &=& \frac{1}{\sqrt{ \epsilon_{\mathrm{sq},+sn}(\bm{k}) }} Q^\dagger(-\bm{k}) U_T \mathcal{K} \ket{u^{s}_{+n}(\bm{k})} \nonumber \\
                         &=& \frac{1}{\sqrt{ \epsilon_{\mathrm{sq},+sn}(\bm{k}) }} Q^\dagger(-\bm{k}) T \ket{u^{s}_{+n}(\bm{k})}  \nonumber \\
                         &=& \mathrm{sgn}(s) \frac{1}{\sqrt{ \epsilon_{\mathrm{sq},+sn}(\bm{k}) }} Q^\dagger(-\bm{k}) \ket{u^{\bar{s}}_{+n}(-\bm{k})}, \nonumber \\
                         &=& \mathrm{sgn}(s) \frac{1}{\sqrt{ \epsilon_{\mathrm{sq},+\bar{s}n}(-\bm{k}) }} Q^\dagger(-\bm{k}) \ket{u^{\bar{s}}_{+n}(-\bm{k})}, \nonumber \\
\end{eqnarray}
where $\bar{s}$ takes $\I$ ($\II$) for $s=\II$ ($\I$).
The function $\mathrm{sgn}(s)$ takes $1$ ($-1$) for $s=\I$ ($\II$). 
From the first and the second line we have used Eq.~(\ref{eq: U_T Q^*(k) U_T=Q(-k)}). 
We note that for time-reversal symmetric $H_{\mathrm{sq},+}(\bm{k})$, the relation $\epsilon_{\mathrm{sq},+\bar{s}n}(-\bm{k})=\epsilon_{\mathrm{sq},+sn}(\bm{k})$ holds.

The above results prove Eq.~(\ref{eq: u- and u+}).

\section{
Derivation of Table~\ref{table: class_table_sqrt-TI}
}
\label{sec: classfication app}

The classification results for systems with the additional symmetry [see Eq.~(\ref{eq: symm on U})] can be obtained by analyzing the symmetry constraints imposed on the block-diagonalized Hamiltonian~\cite{hsq_ftnt}$h_{\mathrm{sq},\pm}(\bm{k})$ with $U$ ($U^2=1$).

\begin{table}[htb]
\begin{center}
\begin{tabular}{ c c c c c c c c } \hline\hline
class & $T$ & $C$ & $\Gamma$ &  $d=0$          & $d=1$           & $d=2$           & $d=3$            \\ \hline
{}$^{\S}$A     &  0  & 0   & 0        &  $\mathbb{Z}$   & 0               & $\mathbb{Z}$    &  0               \\ 
AIII  &  0  & 0   & 1        &  0              & $\mathbb{Z}$    & 0               & $\mathbb{Z}$     \\ \hline
{}$^{\S\S}$AI    &  1  & 0   & 0        &  $\mathbb{Z}$   & 0               & 0               &  0               \\ 
BDI   &  1  & 1   & 1        &  $\mathbb{Z}_2$ &  $\mathbb{Z}$   & 0               &  0               \\ 
D     &  0  & 1   & 0        &  $\mathbb{Z}_2$ &  $\mathbb{Z}_2$ &  $\mathbb{Z}$   &  0               \\ 
DIII  &  -1 & 1   & 1        &  0              &  $\mathbb{Z}_2$ &  $\mathbb{Z}_2$ &  $\mathbb{Z}$    \\ 
{}$^{\S\S\S}$AII   &  -1 & 0   & 0        &  $\mathbb{Z}$   & 0               &  $\mathbb{Z}_2$ &  $\mathbb{Z}_2$  \\ 
CII   &  -1 & -1  & 1        &  0              &  $\mathbb{Z}$   & 0               &  $\mathbb{Z}_2$  \\ 
C     &  0  & -1  & 0        &  0              & 0               &  $\mathbb{Z}$   &  0               \\
CI    & 1   & -1  & 1        &  0              & 0               & 0               &  $\mathbb{Z}_2$  \\  \hline\hline
\end{tabular}
\end{center}
\caption{
Classification results of ordinary topological insulators and superconductors~\cite{Schnyder_classification_free_2008,Kitaev_classification_free_2009,Ryu_classification_free_2010}. 
From the first and the fourth column specify the symmetry of systems. 
The second (third) column describes the presence/absence of the time-reversal (particle-hole) symmetry; for instance, in the second column, ``$\pm1$" indicates presence of time-reversal symmetry [Eq.~(\ref{eq: def of TRS})] described by the time-reversal operator $T^2=\pm1$ while ``$0$" indicates the absence of the symmetry.
In the fourth column, ``$1$" (``$0$") indicates the presence (absence) of the chiral symmetry [Eq.~(\ref{eq: def of CS})].
}
\label{table: class_table_ordinary-TI}
\end{table}

\textit{class A, class AI, and class AII, --.}
In the absence of particle-hole and chiral symmetry, we can take the identity matrix as $U$ satisfying Eq.~(\ref{eq: symm on U}).
Thus, the classification results are identical to the ordinary ones.

\textit{class AIII--.}
Equation~(\ref{eq: symm on U}c) indicates that the chiral symmetry is not closed for the subsectors; no symmetry constraint is imposed on $h_{\mathrm{sq},+}(\bm{k})$.
Thus, $h_{\mathrm{sq},+}(\bm{k})$ belongs to class A. 
Therefore, the classification result is obtained from the row marked with ``$\S$" in Table~\ref{table: class_table_ordinary-TI}.

\textit{class BDI and class CI--.}
Equation~(\ref{eq: symm on U}) indicates that only time-reversal symmetry ($T^2=1$) is closed for each subsector; particle-hole symmetry and chiral symmetry are not closed.
Thus, $h_{\mathrm{sq},+}(\bm{k})$ belongs to class AI. 
Here, we note that the result is not affected by whether $C^2$ is equals to $1$ or $-1$.
Therefore, the classification result is obtained from the row marked with ``$\S\S$" in Table~\ref{table: class_table_ordinary-TI}.

\textit{class DIII and class CII--.}
Equation~(\ref{eq: symm on U}) indicates that only time-reversal symmetry ($T^2=-1$) is closed for each subsector; particle-hole symmetry and chiral symmetry are not closed.
Thus, $h_{\mathrm{sq},+}(\bm{k})$ belongs to class AII. 
As is the previous case, the result is not affected by whether $C^2$ is equals to $1$ or $-1$.
Therefore, the classification result is obtained from the row marked with ``$\S\S\S$" in Table~\ref{table: class_table_ordinary-TI}.

\textit{class D and class C--.}
Equation~(\ref{eq: symm on U}b) indicates that the particle-hole symmetry is not closed for the subsectors; no symmetry constraint is imposed on $h_{\mathrm{sq},+}(\bm{k})$.
Thus, $h_{\mathrm{sq},+}(\bm{k})$ belongs to class A. 
As is the previous case, the result is not affected by whether $C^2$ is equals to $1$ or $-1$.
Therefore, the classification result is obtained from the row marked with ``$\S$" in Table~\ref{table: class_table_ordinary-TI}.

\section{
Topology of the squared Hamiltonian of class AIII in two dimensions
}
\label{sec: AIII 2D app}
The squared Hamiltonian $H_{\mathrm{sq}}(\bm{k})$ can be block-diagonalized with $\Gamma$ ($\Gamma^2=1$).
In this appendix, we show the following facts. (i) The topology of the plus sector is same as the topology of the minus sector. (ii) The presence/absence of the additional symmetry does not affect the topology.

Firstly, we show that the topology of the plus sector is same as the topology of the minus sector [see Eq.~(\ref{eq: NChp =NChm app})]
Because the chiral symmetry is not closed for each subsector, the block-diagonalized Hamiltonian $H_{\mathrm{sq},\pm}$ is characterized by the Chern number
\begin{subequations}
\label{eq: def of Chern}
\begin{eqnarray}
N_{\mathrm{Ch},\pm}&=& \int \frac{dk_xdk_y}{2\pi i} F_{\pm}(\bm{k}),
\end{eqnarray}
with
\begin{eqnarray}
A_{\pm,\mu}(\bm{k})&=& \langle u_{\pm n}(\bm{k})|\partial_\mu |u_{\pm n}(\bm{k})\rangle, \\
F_{\pm}(\bm{k})&=& \partial_x A_{\pm,y}- \partial_y A_{\pm,x}.
\end{eqnarray}
\end{subequations}
Here, $|u_{\pm n}(\bm{k})\rangle$ denote eigenvectors of $H_{\mathrm{sq},\pm}$ with eigenvalues $\epsilon_{\mathrm{sq},+sn}(\bm{k})$.

Now, we show that the relation 
\begin{eqnarray}
\label{eq: NChp =NChm app}
N_{\mathrm{Ch},+} &=& N_{\mathrm{Ch},-},
\end{eqnarray}
holds.

Consider the case where the eigenvalues are non-zero and the Hamiltonian $H(\bm{k})$ is written as
\begin{eqnarray}
H(\bm{k})&=& 
\left(
\begin{array}{cc}
0 & Q(\bm{k}) \\
Q^\dagger(\bm{k}) & 0
\end{array}
\right),
\end{eqnarray}
with $N\times N$-matrix $Q(\bm{k})$ which maps a state in the minus sector of $\Gamma$ to a sate in the plus sector.

As is the case of Eq.~(\ref{eq: u- and u+}), we can see that $|u_{- n}(\bm{k})\rangle$ is obtained from $|u_{+ n}(\bm{k})\rangle$,
\begin{eqnarray}
|u_{- n}(\bm{k})\rangle&=& \frac{1}{\sqrt{ \epsilon_{\mathrm{sq},+n}(\bm{k}) } }Q^\dagger(\bm{k}) |u_{+ n}(\bm{k})\rangle,
\end{eqnarray}
with the eigenvalues of the squared Hamiltonian for the plus sector $\epsilon_{\mathrm{sq},+n}(\bm{k})$.

Thus, we have
\begin{subequations}
\begin{eqnarray}
A_{-,\mu}(\bm{k})&=& A_{+,\mu}(\bm{k}) +f_{\mu}(\bm{k})
\end{eqnarray}
with
\begin{eqnarray}
f_{\mu}(\bm{k}) &=& \frac{1}{\sqrt{ \epsilon_{\mathrm{sq},+n}(\bm{k}) }} \mathrm{tr}[P_+(\bm{k})Q(\bm{k}) \nonumber \\
&& \quad\quad\quad\quad \left(\partial_\mu \frac{1}{\sqrt{ \epsilon_{\mathrm{sq},+n}(\bm{k}) }} Q^\dagger(\bm{k}) \right)   ],\\
P_+(\bm{k}) &=& \sum_{n} \ket{u_{+n}}\bra{u_{+n}}.
\end{eqnarray}
\end{subequations}

Applying Stokes' theorem to Eq.~(\ref{eq: def of Chern}), we can see that the integral of $f_{\mu}(\bm{k})$ vanishes because $f_{\mu}(\bm{k})$ is gauge independent. The above results prove Eq.~(\ref{eq: NChp =NChm app}).

Now, we show that the presence of the additional symmetry $U$ does not change the topology.
For systems with the additional symmetry, applying $U$ just maps a state in the plus sector of $\Gamma$ to a state in the minus sector. 
This fact results in Eq.~(\ref{eq: NChp =NChm app}) which is satisfied even without the additional symmetry.

\section{
Topology of the squared Hamiltonian of class CII in three dimensions
}
\label{sec: CII 3D app}
The squared Hamiltonian $H_{\mathrm{sq}}(\bm{k})$ can be block-diagonalized with $\Gamma$ ($\Gamma^2=1$).
In this appendix, we show the following facts. (i) The topology of the plus sector is same as the topology of the minus sector. (ii) The presence/absence of the additional symmetry does not affect the topology.

Firstly, we show that the topology of the plus sector is same as the topology of the minus sector.
Because only time-reversal symmetry is closed for the block-diagonalized subsector,
the topology of $H_{\mathrm{sq},\pm}(\bm{k})$ is characterized by the $\mathbb{Z}_2$-invariant~\cite{Fu_3DTI_PRL07,Moore_3DTI_PRB07,Roy_3DTI_PRB09} for three-dimensional systems $\nu^{\mathrm{3D}_\pm}$.
For systems whose BZ is cubic, $\nu^{\mathrm{3D}}$ is defined as~\cite{Fu_3DTI_PRL07,Moore_3DTI_PRB07,Fukui_Z2TI_JPSJ07}
\begin{eqnarray}
\label{eq: def of 3D Z2 app}
\nu^{\mathrm{3D}}_{\pm}&=& \nu_{\pm}(0)\nu_{\pm}(\pi).
\end{eqnarray}
Here, $\nu_{\pm}(k^*_z)$ ($k^*_z=0,\pi$) are $\mathbb{Z}_2$-invariants computed for two-dimensional BZs specified by $k^*_z$.
Specifically, $\nu_{\pm}(k^*_z)$ is given by
\begin{eqnarray}
\nu_{\pm}(k^*_z) &=& \frac{1}{2\pi i}\left[ \int^\pi_{-\pi} dk_x [A_{\pm,x}(k_x,\pi,k^*_z)-A_{\pm,x}(k_x,0,k^*_z)]  \right. \nonumber \\
                 &&  \left. -\int^{\pi}_{-\pi} dk_x \int^0_{-\pi} d k_y F_{\pm}(\bm{k}) \right].
\end{eqnarray}
The Berry connection $A_{\pm,\mu}$ ($\mu=x,y$) and the Berry curvature $F(\bm{k})$ are defined in Eq.~(\ref{eq: CII A F}). 
For commutation of $\nu_{\pm}(k^*_z)$, the gauge is chosen so that Eq.~(\ref{eq: gauge constraint}) is satisfied.

Here, we show that 
\begin{eqnarray}
\label{eq: nu3D+=nu3D- app}
\nu^{\mathrm{3D}}_{+} &=&\nu^{\mathrm{3D}}_{-},
\end{eqnarray}
holds, which can be proven in a similar way to the case of two dimensions (see Sec.~\ref{sec: CII Z2 inv proof}~and~Appendix~\ref{sec: proof of u- and u+}).
As is the case of the two-dimensional systems we have
\begin{eqnarray}
\nu_{+}(k^*_z) &=&\nu_{-}(k^*_z),
\end{eqnarray}
without requiring the additional symmetry.
Because the $\nu^{\mathrm{3D}}_{\pm}$ is computed from $\nu_{+}(k^*_z)$ [see Eq.~(\ref{eq: def of 3D Z2 app})], we can see that Eq.~(\ref{eq: nu3D+=nu3D- app}) holds.

Now, we show that the additional symmetry $U$ does not affect the topology.
For systems with the additional symmetry, applying $U$ just maps a state in the plus sector of $\Gamma$ to a state in the minus sector [see Eq.~(\ref{eq: symm on U}c)].
This fact results in Eq.~(\ref{eq: nu3D+=nu3D- app}) which holds even without the additional symmetry.

\end{document}